# Flux jumps at pulsed field magnetization

V. S. Korotkov [1], E. P. Krasnoperov [1,2], A. A. Kartamyshev [1]

[1] NRC "Kurchatov Institute", Russia, 123182, Moscow
[2] MIPT, Russia, 141700, Dolgoprudniy

**Abstract.** Flux jumps lead to 10-20 times reduction of the shielding currents at a pulsed field magnetization of a melt-grown HTS annuli. The induced circular currents and the related trapped field are small after a field pulse since the duration of the pulse is shorter than the temperature relaxation time (<<1 s). Moreover, the residual trapped field in the hole of the annuli has the opposite direction in respect to the applied field. It is shown that the small value of the induced circular currents at the relatively high average critical current density is related to a narrow angular region of the flux motion ( $\approx 7^0$).

### 1. Introduction

Flux jumps in hard superconductors are well studied. They have avalanche-like nature and can take place both at low perturbations and at a high rate of the magnetic field variation [1]. The flux jumps may occur due to thermomagnetic instability at low temperatures in conditions of the slow magnetization (close to the isothermal magnetization). Accidental thermal perturbations cause the motion of the flux structure inducing the electric field. This process leads to the energy dissipation and the temperature rise. The critical current decreases if the heating stimulated by the flux motion exceeds the initial heat perturbation. This, in turn, leads to the avalanche process – the flux jump. However, the thermomagnetic instability is not observed in the HTS at the temperatures above 40 K due to the large heat capacity [2].

Flux jumps at high temperatures can occur if perturbations is large enough, in particular, at high rate of field variation (field ramp) [1]. To the best of our knowledge, there are no data in the literature about the flux jumps in spite of the numerous pulse field magnetization (PFM) experiments on HTS melt-grown discs. However, the PFM is important for applications as economical and autonomous method of strong magnetic fields generation. Therefore, the lack of these relevant data on the pulsed field magnetization of annuli stimulated our research. In this work the PFM dynamics in melt-grown HTS annuli was studied at the liquid nitrogen temperature (T = 78K). The evolution of the circular currents and the trapped magnetic fields were measured. The distribution of the trapped field and the critical current density were determined at different magnetization levels.

### 2. Experimental setup

Four HTS annuli with outer diameter $2a_1 = 36$ mm, inner diameter $2a_0 = 17$ mm and thickness $2b \approx 11$ mm were investigated. The annuli were cut from melt - grown Y-Ba-Cu-O discs [3]. The superconducting critical temperatures of the annuli were approximately about 90K and the critical current densities were differed by 10-20 %, only. However, after the cycles of cooling-heating some unprotected samples have deteriorated. As a result, the maximum trapped fields were reduced by almost 2 times (2-3 years). The described effects were identical in all studied annuli. Therefore, the experimental data of two representative annuli are shown below. The maximum trapped fields were $B_{tr} \approx 0.4$ T at T = 78 K. The magnetizing field was generated by copper coils located outside annuli. The current source allowed generating pulses with a peak field up to 3.5 T and the duration up to 40 ms. The magnetic field was measured by the Hall probe which was connected in parallel to the digital oscilloscope and to the DC micro-voltmeter. Such way we could measure the field evolution during the pulse and a field value after the magnetization with high accuracy. For the measurement of the circular shielding current the Rogowski belt (Ø 25 mm, the cross section 4x1.5 mm) was used.

### 3. Results and Discussion

The dependence of the trapped field in the center of the annuli versus pulsed field amplitude is shown in fig. 1. After each step of the magnetization the annuli were heated above $T_c$. At low pulse field amplitudes ($\mu_0 H_a < 0.5$ T) the trapped field is zero ($B_{tr} \approx 0$). At these conditions, the circular currents screen completely the external field. The $B_{tr}$ appears and increases quickly with increasing the pulse field amplitude above 0.5 T. At some critical value $\mu_0 H_a \approx 1.0$ T the trapped field has a maximum. With further increase of the $\mu_0 H_a$ the trapped field drops abruptly and changes the direction (sign). The sharp drop of $B_{tr}$ and the reversing its direction is caused by a flux jump. Obviously, such strong field drop happens as a result of partial or full destruction of shielding current circulated around the hole.

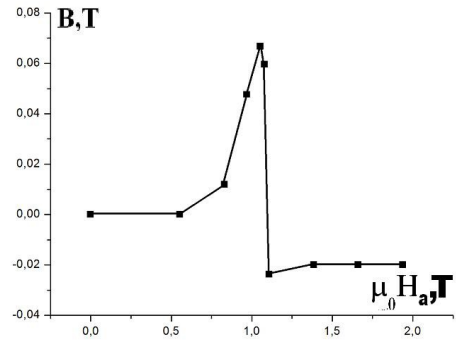

Fig. 1. The trapped field in the center of the annuli as a function of the pulse field amplitude.

In case of slow isothermal magnetization of the thick-walled cylinder with internal diameter $2a_0$ and external diameter $2a_1$, the highest trapped field according to Bean model is $B_m = \mu_0 J_C (a_1 - a_0)$. To reach this value, one should apply the external field equal or exceeding $2B_m$ [4, 5]. At PFM of annuli, taking into account the demagnetization and the magnetic field diffusion, the higher amplitude of the field is required. However, flux jumps do not allow to achieve the maximal value of the trapped field at a pulsed field magnetization. They arise due to rapid variation of the external magnetic field. For the maximal magnetization ($B_m = 0.4$T) the multi- pulse process (series of pulses) with gradual amplitude increase is required [6,7]. In this regime the flux jumps do not occur.

The evolution of the shielding current and the magnetic fields during flux jumps are presented in fig. 2. The right axis relates to the external field of $\mu_0 H_a \approx 2$ T (dots) and the field in the center of the annuli (solid curve). The left axis relates to the shielding current evolution curve I(t). Up to $t=t_C$ the current rises and then drops abruptly in 10 times. The field induced by shielding current is equal to the external field within 5-7 % on the increasing part of the current curve. After the sharp drop of the shielding current the flux penetrates into the hole of annuli. This fast field penetration is the flux jump. The Hall probe shows that starting with this moment of time the field in the center becomes nearly the same as the external field. The wavy curve of the Hall sensor data in Fig. 2 is a result of averaging procedure of the signal, which is necessary due to the low sensitivity of the pulsed oscilloscope (1-2 mV).

The current drop is well described by the exponential function $I=I_C \exp(-t/\tau_f)$, where $\tau_f=L/R_{flow}$, (L – the inductance, $R_{flow}$ – the flux flow resistance). The $I_C$ is the critical current. The characteristic time of the current drop is $\tau_f = 0.36$ ms. This time is independent on the magnetic field up to 5 T and the pulse duration down to 10 ms. In order to determine the $R_{flow}$, the damping time of the current was measured in pure aluminum annuli having identical sizes. It turned out that for Al (at T = 78 K) the damping time is order of magnitude higher and is $\tau_{Al} = 3.6$ ms. The estimated resistance of the Al annuli is $R_{al}=2\pi \cdot \rho_{Al}(b \cdot \ln(a_1/a_0))^{-1}=$ 2.5μΩ, using the resistivity of Al at T = 78 K is $\rho_{Al}$ = 0.2-0.3μΩ·cm [5]. Accordingly, the flux flow resistance in the superconductor is $R_{flow} = R_{Al} \cdot \tau_{Al} / \tau_f$ =25 μΩ.

The shielding currents remain small when the external field decreases. Therefore, the field in the hole is close to the external field. The low currents after the flux jump and the slow rate of their variation are explained by the high temperature of annuli, which is a consequence of the high heat capacity and the low heat conductivity of HTS. The corresponding thermal relaxation time in our samples is $\tau_{th} \approx$ 1s [6]. As a result the flux dynamics remain high and the difference between the external and the internal field is practically absent after flux jump.

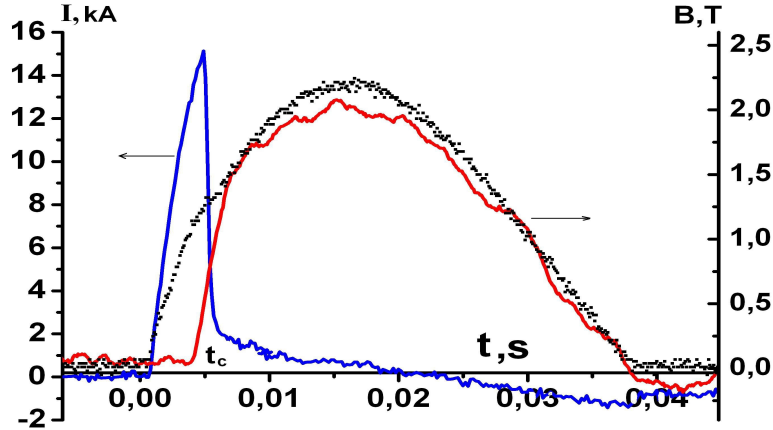

Fig. 2. The evolution of the current and the field. Left axis – the current in annuli – I(t). Right axis – the external field (dots) and the field in the hole (solid curve).

It is well known that the destruction of the critical current and corresponding flux jump are caused by the overheating. In the experiment shown in Fig.2 the critical current drops in 10 times. For such strong reduction of the current the heating on $\Delta T_j \approx$10 K is required, assuming the linear temperature approximation of current I(T). On the other hand the dissipated heat during the jump is $q_j$ = ∫ $I^2 \cdot R_{flow}$ dt ≈ 1.28 J. The corresponding average temperature rise <ΔT>= $q_j$/ $C_\Omega \cdot \Omega$ (here $C_\Omega$ - volumetric heat capacity [7]) is less than 0.2 K if this heat is dissipated into the whole volume of the annulus. Such low heating is insufficient to explain the observed current drop during the flux jump.

To find out the distribution of the current in the superconducting annuli the radial trapped field distribution $B_z(r)$ were measured in the central plane z = 0. Two identical annuli with 2 mm gap were used for the measurements. The axis component of the field was measured by the Hall probe which was moved along the radial direction. The current density was estimated using the expression $J=\mu_0^{-1}dB_z/dr$. $B_z(r)$ in the gap between annuli for the different magnetization conditions is shown in fig. 3. The curve (1) corresponds to the single pulse magnetization with the field amplitude of $\mu_0H_a$= 0.9 T. The curve (2) corresponds to the multi- pulse magnetization (10 pulses). The curve (3) show the field distribution after the flux jump at the applied field $\mu_0H_a$ =2.0 T.

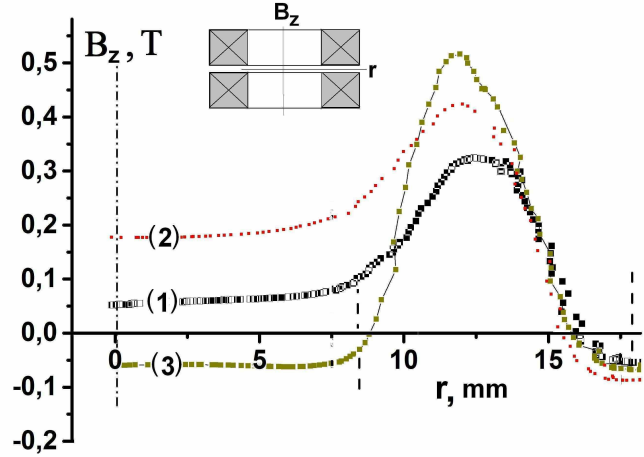

Fig. 3. The radial field distribution $B_z(r)$ at the different conditions: (1) – after the single pulse of $\mu_0 H_a = 0.9$ T; (2) – after 10 pulses with peak field $\mu_0 H_a = 0.9$ T; (3) after the flux jump $\mu_0 H_a = 2.0$ T. The vertical dotted lines corresponds to the inner and outer annuli radii.

Currents in the region of $dB_z/dr < 0$ produces the field in the direction of the magnetizing field while the internal currents ($dB_z/dr > 0$) produce the field with the opposite direction. As shown in fig.1 when the current drops the magnetic flux penetrates into the hole. At a pulse field magnetization, one would expect that the flux trapping can be observed if $I_c$ is high. Otherwise, (if $I_c$ is small) the flux escapes the hole and low field will be trapped. In the experiment, the trapped field is low and has the opposite direction respect to the applied field. On the other hand from the fig. 3 it follows than the average critical current density $J_c = \mu_0^{-1} dB_z/dr = 15$ kA/cm$^2$ is high enough. The high critical current density corresponds to the relatively low (<0.5 K) heating as it is observed in the experiments [6,8]. Therefore, we assume that flux jumps lead to a local heat dissipation and, correspondingly, suppress the critical current density only locally in the region of low pinning force. As it was estimated above, the local heating should be at least $\Delta T_j \approx 10$K. The volume of the region occupied by the moving fluxoid at the flux jump can be estimated from the relation

$$\Omega_{flow} = q_j / C_\Omega \cdot \Delta T_j$$

Due to the rotational symmetry, it is reasonable to assume that this region corresponds to the sector of annulus. Thus we can obtain that sector angle is of $\varphi = 2\pi \cdot \langle \Delta T \rangle / \Delta T_j \approx 7^0$.

The local heat dissipation in the narrow sector during the flux jump is explained by the azimuthal inhomogeneity of the critical current. In the melt-grown Y-2-3 the critical currents in (ab) plane have angle anisotropy which manifests itself in the anisotropy of the trapped magnetic field in discs [8]. This inhomogeneity of the critical current can be the triggered by the technology of the HTS growth. Even if the anisotropy is low (10-15%) due to the high exponent of HTS current-voltage curve ($E \sim J_c^N$, где N>25 [9]), the temperature of the soft flux region during the field pulse exceeds the temperature of region with higher $J_c$. Thus, the most of energy will be dissipated in the narrow sector of annuli having the lowest local $J_c$.

Using the volume of soft flux region we can estimate the dynamical resistance of the flux motion. For this narrow sector of the annuli, we can write:

$$\rho_{flow} = (2\pi/\varphi) \cdot \rho_{Al} \cdot (\tau_{Al} / \tau_f) \approx 100 \, \mu\Omega \cdot cm$$

This value is approximately two time smaller than the resistivity in the ab-plane at the normal state

[11]. The region of flux motion remains at the high temperature after the flux jump due to the short pulse duration compared to the thermal relaxation time. The rest of the annuli temperature is lower and Jc maintains its high value. This experimental situation is similar to the case of annuli with a slit. Therefore, when the external field is switched off, the residual magnetization has a maximum in the body of the annuli and low negative value near the internal and external edge of it. The symmetrical configuration of the magnetic field shown in fig 3 creates the negative value in the center of the annulus when the radius of the annuli is comparable to its thickness. Similar distribution of $B_z$ (r) is observed in the annuli with a slit when magnetic field is removed after cooling in the static magnetic field [15].

### 4. Conclusion

We have shown that the circular currents are destructed and the flux jump takes place at the pulsed field magnetization of melt-grown HTS annuli. The vortexes retain high mobility due to the slow thermal processes. Therefore, the magnetic flux is not trapped in the hole of annuli. The low value of the critical current and high value of the critical current density in the body of annuli indicate that: 1) during the flux jump the magnetic field penetrates into the hole through the soft flux region of the annulus; 2) the soft flux region is the narrow sector of the annuli where the heating during the field pulse at least in order of magnitude higher than the average heating.

### Acknowledgments.

Authors are grateful to N.A. Nizhelsiy for the melt grown annuli fabrication and to D.S. Yashin for Rogowsky coil fabrication. One of authors V.S.K. appreciate for the financial support from the Foundation of Assistance Small Entrepreneurship in Science (FASIE). Special thanks V.Grinenko for criticism and proofreading text.